\documentclass[]{acm_proc_article-sp}
\usepackage{amsmath,epsfig,latexsym,psfig,url,graphicx}

\renewcommand{\footnotesize}{\small}

\newtheorem{theorem}{Theorem}
\newtheorem{lemma}{Lemma}

\def\up{\vspace*{-.4cm}}

\def\compactify{\itemsep=0pt \topsep=0pt \partopsep=0pt \parsep=0pt}
 \let\latexusecounter=\usecounter

\title{A Price-Anticipating Resource Allocation Mechanism for
Distributed Shared Clusters}

\newcount\aucount
\newcount\originalaucount
\newdimen\auwidth
\newdimen\auskip
\newcount\auskipcount
\newdimen\auskip
\global\auskip=1pc
\newdimen\allauboxes
\allauboxes=\auwidth
\newtoks\addauthors
\newcount\addauflag
\global\addauflag=0 

\gdef\numberofauthors#1{\global\aucount=#1
\ifnum\aucount>4\global\originalaucount=\aucount \global\aucount=4\fi 
\global\auskipcount=\aucount\global\advance\auskipcount by 1
\global\multiply\auskipcount by 2
\global\multiply\auskip by \auskipcount
\global\advance\auwidth by -\auskip
\global\divide\auwidth by \aucount}

\newfont{\affname}{phvr at 10pt}
\newfont{\eaddrfnt}{phvr at 8pt}
\newfont{\authorfont}{phvr at 12pt}
\def\email#1{{{\eaddrfnt{\vskip 4pt#1}}}}

\def\alignauthor{
\end{tabular}%
  \begin{tabular}[t]{p{\auwidth}}\centering}%

\numberofauthors{4}

\author{
  \alignauthor \authorfont{Michal Feldman}\thanks{School of Information Management and System, University of California, Berkeley, CA 94720}
  \email{mfeldman@sims.berkeley.edu}\\
      \alignauthor \authorfont{Kevin Lai}\thanks{Information Dynamics Laboratory, HP Labs, Palo Alto, CA 94304}
      \email{kevin.lai@hp.com}\\
      \alignauthor \authorfont{Li Zhang}$^\dagger$
      \email{l.zhang@hp.com}\\
          }

\date{}

\begin{document}




\maketitle


\begin{abstract}

\small 


In this paper we formulate the fixed budget resource allocation game
to understand the performance of a distributed market-based resource
allocation system. Multiple users decide how to distribute their
budget (\emph{bids}) among multiple machines according to their
\emph{individual preferences} to maximize their individual utility. We
look at both the efficiency and the fairness of the allocation at the
equilibrium, where fairness is evaluated through the measures of
\emph{utility uniformity} and \emph{envy-freeness}. We show
analytically and through simulations that despite being highly
decentralized, such a system converges quickly to an equilibrium and
unlike the \emph{social optimum} that achieves high efficiency but
poor fairness, the proposed allocation scheme achieves a nice balance
of high degrees of efficiency and fairness at the equilibrium.
\end{abstract}


\section{Introduction}

The primary advantage of distributed shared clusters like the Grid
\cite{foster1997} and PlanetLab \cite{planetlab2003} is their ability
to pool together shared computational resources. This allows increased
throughput because of statistical multiplexing and the bursty
utilization pattern of typical users. Sharing nodes that are dispersed
in the network allows lower delay because applications can store data
close to users.  Finally, sharing allows greater reliability because
of redundancy in hosts and network connections.

However, resource allocation in these systems remains an issue.  The
problem is how to allocate a shared resource both fairly and
efficiently (where efficiency is the ratio of the achieved social
welfare to the optimal one). The main challenge is \emph{strategic}
users who act in their own interests.

Several non-economic allocation algorithms have been proposed, but
these typically assume that task values (i.e., their
importance) are the same, or are inversely proportional to the
resources required, or are set by an omniscient administrator.
However, in many cases, task values vary significantly, are not
correlated to resource requirements, and are difficult and
time-consuming for an administrator to set. Instead, we examine a
market-based resource allocation system (others are described
in~\cite{ferguson1988, waldspurger1992, regev1998, chun2000,
wellman2001, auyoung2004}) that allows users to express their
preferences for resources through a bidding mechanism.

In particular, we consider a \emph{price-anticipating} scheme in which
a user bids for a resource and receives the ratio of his
bid to the sum of bids for that resource.  This proportional scheme is
simpler, more scalable, and more responsive~\cite{lai2004-2} than
auction-based schemes~\cite{ferguson1988, waldspurger1992, regev1998}.
Previous work has analyzed price-anticipating schemes in the context
of allocating network capacity for flows for users with unlimited
budgets. In this work, we examine a price-anticipating scheme in the
context of allocation computational capacity for users with private
preferences and limited budgets, resulting in a qualitatively
different game (as discussed in Section~\ref{sec:relatedwork}).

In this paper, we formulate the \emph{fixed budget} resource
allocation game and study the existence and performance of the Nash
equilibria of this game.  For evaluating the Nash equilibria, we
consider both their efficiency, measuring how close the social welfare
at equilibrium is to the social optimum, and fairness, measuring how
different the users' utilities are. Although rarely considered in
previous game theoretical study, we believe fairness is a critical
metric for a resource allocation schemes because the perception of
unfairness will cause some users to reject a system with more
efficient, but less fair resource allocation in favor of one with less
efficient, more fair resource allocation. We use both \emph{utility
uniformity} and \emph{envy-freeness} to measure fairness. Utility
uniformity, which is common in Computer Science work, measures the
closeness of utilities of different users. Envy-freeness, which is
more from the Economic perspective, measures the happiness of users
with their own resources compared to the resources of others.

Our contributions are as follows:


{$\bullet$ \bf We analyze the existence and performance of Nash
equilibria.} Using analysis, we show that that there is always a Nash
equilibrium in the fixed budget game if the utility functions satisfy
a fairly weak and natural condition of \emph{strong competitiveness}.
We also show the worst case performance bounds: for $m$ players the efficiency at
equilibrium is $\Omega(1/\sqrt{m})$, the utility uniformity is $\geq
1/m$, and the envy-freeness $\geq 2\sqrt{2}-2 \approx 0.83$. Although
these bounds are quite low, the simulations described below indicate
these bounds are overly pessimistic.

{$\bullet$ \bf We describe algorithms that allow strategic users to
optimize their utility.} As part of the fixed budget game analysis, we
show that strategic users with linear utility functions can calculate
their bids using a \emph{best response} algorithm that quickly results
in an allocation with high efficiency with little
computational and communication overhead. We present variations of
the best response algorithm for both finite and infinite parallelism
tasks. In addition, we present a \emph{local greedy adjustment}
algorithm that converges more slowly than best response, but allows
for non-linear or unformulatable utility functions.


{$\bullet$ \bf We show that the price-anticipating resource allocation
mechanism achieves a high degree of efficiency and fairness.} Using
simulation, we find that although the socially optimal allocation
results in perfect efficiency, it also results in very poor
fairness. Likewise, allocating according to only users' preference
weights results in a high fairness, but a mediocre
efficiency. Intuition would suggest that efficiency and fairness are
exclusive. Surprisingly, the Nash equilibrium, reached by each user
iteratively applying the best response algorithm to adapt his bids,
achieves nearly the efficiency of the social optimum and nearly the
fairness of the weight-proportional allocation: the efficiency is
$\geq 0.90$, the utility uniformity is $\geq 0.65$, and the
envy-freeness is $\geq 0.97$, independent of the number of users in
the system. In addition, the time to converge to the equilibrium is
$\leq 5$ iterations when all users use the best response strategy. The
local adjustment algorithm performs similarly when there is sufficient
competitiveness, but takes $25$ to $90$ iterations to stabilize.


As a result, we believe that shared distributed systems based on the
fixed budget game can be highly decentralized, yet achieve a high
degree of efficiency and fairness.

The rest of the paper is organized as follows. We describe the model
in Section~\ref{sec:model} and derive the performance at the Nash
equilibria for the infinite parallelism model in Section~\ref{sec:results}. In
Section~\ref{sec:algorithms}, we describe algorithms for users to
optimize their own utility in the fixed budget game. In
Section~\ref{sec:simulation_results}, we describe our simulator and
simulation results. We describe related work in
Section~\ref{sec:relatedwork} and conclude in
Section~\ref{sec:conclusions}. 


\def\note#1{\textbf{#1}}
\def\vecr{\textbf{{\textrm{r}}}}
\def\vecx{\textbf{{\textrm{x}}}}
\def\vecs{\textbf{{\textrm{s}}}}
\def\vecp{\textbf{{\textrm{p}}}}
\def\vecq{\textbf{{\textrm{q}}}}

\section{The model}
\label{sec:model}

In this section, we formally describe the model we work with
and define the terms and notations throughout the paper.

\textbf{Price-Anticipating Resource Allocation.}\hspace*{.2cm}
We study the problem of allocating a set of divisible resources (or
machines). Suppose that there are $m$ users and $n$ machines. Each
machine can be continuously divided for allocation to multiple users.
An \emph{allocation scheme} $\omega = (\vecr_1, \ldots, \vecr_m)$,
where $\vecr_i=(r_{i1},\cdots,r_{in})$ with $r_{ij}$ representing the
share of machine $j$ allocated to user $i$, satisfies that for any
$1\leq i\leq m$ and $1\leq j\leq n$, $r_{ij}\geq 0$ and $\sum_{i=1}^m
r_{ij}\leq 1$.  Let $\Omega$ denote the set of all the allocation
schemes.

We consider the \emph{price anticipating mechanism} in which each user
places a bid to each machine, and the price of the machine is
determined by the total bids placed.  Formally, suppose that user $i$
submits a non-negative bid $x_{ij}$ to machine $j$.  The price of
machine $j$ is then set to $Y_j=\sum_{i=1}^n x_{ij}$, the total bids
placed on the machine $j$.  Consequently, user $i$ receives a
fraction of $r_{ij}=\frac{x_{ij}}{Y_j}$ of $j$. When $Y_j=0$,
i.e. when there is no bid on a machine, the machine is not allocated
to anyone.  We call $\vecx_i=(x_{i1},\ldots, x_{in})$ the bidding
vector of user $i$.


The additional consideration we have is that each user $i$ has a
budget constraint $X_i$.  Therefore, user $i$'s total bids have to
sum up to his budget, i.e.\ $\sum_{j=1}^n x_{ij} = X_i$.  The budget
constraints come from the fact that the users do not have infinite
budget. 

\textbf{Utility Functions.}\hspace*{.2cm}
Each user $i$'s utility is represented by a function $U_i$ of the
fraction $(r_{i1},\ldots, r_{in})$ the user receives from each
machine.  Given the problem domain we consider, we assume that each
user has different and relatively independent preferences for different machines.
Therefore, the basic utility function we
consider is the \emph{linear utility function}: $U_i(r_{i1},\cdots,
r_{in}) = w_{i1}r_{i1}+\cdots+w_{in}r_{in}$, where $w_{ij}\geq 0$ is
user $i$'s private preference, also called his \emph{weight}, on
machine $j$.  For example, suppose machine $1$ has a faster CPU but
less memory than machine $2$, and user $1$ runs CPU bounded
applications, while user $2$ runs memory bounded applications. As a
result, $w_{11}>w_{12}$ and $w_{21}<w_{22}$.  

Our definition of utility functions corresponds to the user having
enough jobs or enough parallelism within jobs to utilize all the
machines. Consequently, the user's goal is to grab as much of a
resource as possible.  We call this the \emph{infinite parallelism
model}. In practice, a user's application may have an inherent limit
on parallelization (e.g., some computations must be done sequentially)
or there may be a system limit (e.g., the applications data is being
served from a file server with limited capacity).  To model this, we
also consider the more realistic \emph{finite parallelism} model,
where the user's parallelism is bounded by $k_i$, and the user's
utility $U_i$ is defined to be the sum of $k_i$ largest
$w_{ij}r_{ij}$. In this model, the user only submits bids to up to
$k_i$ machines.

There are other system issues that this model does not consider. There
is usually a delay before a user can utilize a machine. For example,
the user must copy code and data to a machine before running his
application there. This model assumes that the job size is large
enough that this delay is negligible. Another issue is that there is
overhead for multiplexing resources on a single machine. This model
assumes the degree of multiplexing is sufficiently low that the
overhead of modern operating system and virtualization technologies is
negligible. 

\textbf{Best Response.}\hspace*{.2cm} 
As typically, we assume the users are selfish and strategic --- they
all act to maximize their own utility, defined by their utility
functions.  From the perspective of user $i$, if the total bids of the
other users placed on each machine $j$ is $y_j$, then the \emph{best
response} of user $i$ to the system is the solution of the following
optimization problem:
\[\mbox{maximize $U_i(\frac{x_{ij}}{x_{ij}+y_j})$ subject to}\]
\[\mbox{$\sum_{j=1}^n x_{ij} = X_i$, and $x_{ij}\geq 0$.}\]

The difficulty of the above optimization problem depends on the
formulation of $U_i$.  We will show later how to solve it for the
infinite parallelism model and provide a heuristic for finite
parallelism model.

\textbf{Nash Equilibrium.}\hspace*{.2cm}
By the assumption that the user is selfish, each user's bidding vector
is the best response to the system.  The question we are most
interested in is whether there exists a collection of bidding vectors,
one for each user, such that each user's bidding vector is the best
response to those of the other users.  Such a state is known as the \emph{Nash
equilibrium}, a central concept in Game Theory.  Formally, the bidding
vectors $\vecx_1,\ldots,\vecx_m$ is a \emph{Nash equilibrium} if for
any $1\leq i\leq m$, $\vecx_i$ is the best response to the system, or,
for any other bidding vector $\vecx_i'$,
\[U_i(\vecx_1,\ldots, \vecx_i, \ldots,\vecx_m) \geq U_i(\vecx_1,
\ldots,\vecx_i',\ldots, \vecx_m)\,.\]

The Nash equilibrium is desirable because it is a stable state at
which no one has incentive to change his strategy.  But a game may not
have an equilibrium.  Indeed, Nash equilibrium may not exist in
the price anticipating scheme we defined above.  This can be shown by
a simple example of two players and two machines.  For example, let
$U_1(r_1, r_2)=r_1$ and $U_2(r_1,r_2)=r_1+r_2$.  Then player $1$
should never bid on machine $2$ because it has no value to him.  Now,
player $2$ has to put a positive bid on machine $2$ in order to claim
the machine, but there is no low limit, resulting the non-existence
of the Nash equilibrium.  Clearly, this happens whenever there is a
resource that is ``wanted'' by only one player.  To rule out this
case, we consider those \emph{strongly competitive} games.  Under the
infinite parallelism model, a game is called strongly competitive if
for any $1\leq j\leq n$, there exist $i\neq k$ such that
$w_{ij},w_{kj}>0$.  Under such condition, we have that
\begin{theorem}\label{thm:exist}
There always exists a pure strategy Nash equilibrium in a strongly
competitive game. 
\end{theorem}
\begin{proof}
The result is a special case of the general result in~\cite{zhang2004}.
We have included in Appendix~\ref{sec:p1} a simpler proof for linear
utility functions considered in this paper.
\end{proof}

\up
Given the existence of the Nash equilibrium, the next important
question is the performance at the Nash equilibrium, which is often
measured by its efficiency and fairness.

\textbf{Efficiency (Price of Anarchy).}\hspace*{.2cm}
For an allocation scheme $\omega\in\Omega$, denote by
$U(\omega)=\sum_i U_i(\vecr_i)$ the social welfare under $\omega$.
Let $U^\ast = \max_{\omega\in\Omega} U(\omega)$ denote the optimal
social welfare --- the maximum possible aggregated user utilities.
The efficiency at an allocation scheme $\omega$ is defined as
$\pi(\omega)=\frac{U(\omega)}{U^\ast}$.  Let $\Omega_0$ denote the set
of the allocation at the Nash equilibrium.  When there exists Nash
equilibrium, i.e.\ $\Omega_0\neq\emptyset$, define the
\emph{efficiency} of a game $Q$ to be $\pi(Q)=\min_{\omega\in\Omega_0}\pi(\omega)$.

It is usually the case that $\pi<1$, i.e. there is an efficiency loss
at a Nash equilibrium.  Such loss can be viewed as caused by the lack
of central enforcement of the user's behavior.  Thus, such efficiency
loss is also called the \emph{price of anarchy}, a term coined
in~\cite{papadimitriou2001}.  The price of anarchy is an important
metric as it represents the maximum social loss as a result of the
users' rationality and the decentralization of the allocation scheme.

\textbf{Fairness.}\hspace*{.2cm}
While the definition of efficiency is standard, there are multiple
ways to define fairness.  We consider two metrics.  One is by
comparing the users' utilities.  The \emph{utility uniformity}
$\tau(\omega)$ of an allocation scheme $\omega$ is defined to be
$\frac{\min_i U_i(\omega)}{\max_i U_i(\omega)}$, the ratio of the
minimum utility and the maximum utility among the users.  Such
definition (or utility discrepancy defined similarly as $\frac{\max_i
U_i(\omega)}{\min_i U_i(\omega)}$) is used extensively in Computer
Science literature.  Under this definition, the utility uniformity
$\tau(Q)$ of a game $Q$ is defined to be $\tau(Q)=\min_{\omega\in\Omega_0}\tau(\omega)$.

The other metric extensively studied in Economics is the concept of
envy-freeness~\cite{varian1974}.  Unlike the utility uniformity metric,
the enviness concerns how the user perceives the value of the share
assigned to him, compared to the shares other users receive.  Within
such a framework, define the \emph{envy-freeness} of an allocation
scheme $\omega$ by $\rho(\omega)=\min_{i,j}\frac{U_i(\vecr_i)}{U_i(\vecr_j)}$.  When $\rho(\omega)\geq 1$, the
scheme is known as an envy-free allocation scheme.  Likewise, the
envy-freeness $\rho(Q)$ of a game $Q$ is defined to be $\rho(Q)=\min_{\omega\in\Omega_0}\rho(\omega)$.

\def\set#1{\{#1\}}
\long\def\commented#1{{}}
\section{Nash Equilibrium}\label{sec:results}

In this section, we present some theoretical results regarding the
performance at Nash equilibrium under the infinite parallelism model.
We assume that the game is strongly competitive to guarantee the
existence of equilibria.  For a meaningful discussion of efficiency
and fairness, we assume that the users are symmetric by requiring that
$X_i=1$ and $\sum_{j=1}^n w_{ij}=1$ for all the $1\leq i\leq m$.  Or
informally, we require all the users have the same budget, and they
have the same utility when they own all the resources.  This precludes
the case when a user has an extremely high budget, resulting in very
low efficiency or low fairness at equilibrium.

We first provide a characterization of the equilibria.  By definition,
the bidding vectors $\vecx_1,\ldots,\vecx_m$ is a Nash equilibrium if
and only if each player's strategy is the best response to the group's
bids.  Since $U_i$ is a linear function and the domain of each users
bids $\set{(x_{i1},\ldots,x_{in})|\sum_j x_{ij}=X_i\,,\mbox{and $x_{ij}\geq 0$}}$
is a convex set, the optimality condition is that there exists
$\lambda_i>0$ such that
\begin{equation}\label{eqn:opt}
\frac{\partial U_i}{\partial x_{ij}} = w_{ij}\frac{Y_j-x_{ij}}{Y_j^2}
\left\{
\begin{array}{lll}
= & \lambda_i & \mbox{if $x_{ij}>0$, and}\\
< & \lambda_i & \mbox{if $x_{ij}=0$.}
\end{array}
\right.
\end{equation}

Or intuitively, at an equilibrium, each user has the same marginal
value on machines where they place positive bids and has lower marginal
values on those machines where they do not bid.

Under the infinite parallelism model, it is easy to compute the social
optimum $U^\ast$ as it is achieved when we allocate each machine
wholly to the person who has the maximum weight on the machine,
i.e. $U^\ast = \sum_{j=1}^n \max_{1\leq i\leq m} w_{ij}$.

\subsection{Two-player Games}

We first show that even in the simplest nontrivial case when there are
two users and two machines, the game has interesting properties.  We
start with two special cases to provide some intuition about the game.
The weight matrices are shown in figure~\ref{fig:two_player_games}(a)
and (b), which correspond respectively to the equal-weight and
opposite-weight games.  Let $x$ and $y$ denote the respective bids of
users $1$ and $2$ on machine $1$.  Denote by $s=x+y$ and
$\delta=(2-s)/s$.

\textbf{Equal-weight game.}\hspace*{.2cm}
In Figure~\ref{fig:two_player_games}, both users have equal valuations
for the two machines. By the optimality condition, for the bid vectors
to be in equilibrium, they need to satisfy the following equations
according to~(\ref{eqn:opt})
\begin{eqnarray*}
\alpha\frac{y}{(x+y)^2} & = &(1-\alpha)\frac{1-y}{(2-x-y)^2}\\
\alpha\frac{x}{(x+y)^2} & = &(1-\alpha)\frac{1-x}{(2-x-y)^2}
\end{eqnarray*}

By simplifying the above equations, we obtain that $\delta =
1-1/\alpha$ and $x=y=\alpha$.  Thus, there exists a unique Nash
equilibrium of the game where the two users have the same bidding
vector.  At the equilibrium, the utility of each user is $1/2$, and
the social welfare is $1$.  On the other hand, the social optimum is
clearly $1$.  Thus, the equal-weight game is ideal as the efficiency,
utility uniformity, and the envy-freeness are all $1$.
\begin{figure}
\begin{center}
\epsfig{file=figures/two_player_games.epsi,width=2.5in}
\caption{\small Two special cases of two-player games.}
\label{fig:two_player_games}
\end{center}
\end{figure}

\textbf{Opposite-weight game.}\hspace*{.2cm}
The situation is different for the opposite game in which the two
users put the exact opposite weights on the two machines.  Assume
that $\alpha\geq 1/2$.  Similarly, for the bid vectors to be at the
equilibrium, they need to satisfy
\begin{eqnarray*}
\alpha\frac{y}{(x+y)^2} & = &(1-\alpha)\frac{1-y}{(2-x-y)^2}\\
(1-\alpha)\frac{x}{(x+y)^2} & = & \alpha\frac{1-x}{(2-x-y)^2}
\end{eqnarray*}

By simplifying the above equations, we have that each Nash equilibrium
corresponds to a nonnegative root of the cubic equation
$f(\delta) = \delta^3 - c\delta^2+c\delta - 1 = 0$, 
where $c=\frac{1}{2\alpha(1-\alpha)}-1$.  

Clearly, $\delta=1$ is a root of $f(\delta)$.  When $\delta=1$, we have
that $x=\alpha$, $y=1-\alpha$, which is the symmetric equilibrium that
is consistent with our intuition --- each user puts a bid proportional
to his preference of the machine.  At this equilibrium,
$U=2-4\alpha(1-\alpha)$, $U^\ast = 2\alpha$, and
$U/U^\ast=(2\alpha+\frac{1}{\alpha})-2$, which is minimized when
$\alpha=\frac{\sqrt{2}}{2}$ with the minimum value of
$2\sqrt{2}-2\approx 0.828$.  
However, when $\alpha$ is large enough, there exist two other roots,
corresponding to less intuitive asymmetric equilibria.

Intuitively, the asymmetric equilibrium arises when user $1$ values
machine $1$ a lot, but by placing even a relatively small bid on
machine $1$, he can get most of the machine because user $2$
values machine $1$ very little, and thus places an even smaller bid. In
this case, user $1$ gets most of machine $1$ and almost half of
machine $2$.  




The threshold is at when $f'(1)=0$, i.e.\ when
$c=\frac{1}{2\alpha(1-\alpha)}=4$.  This solves to
$\alpha_0=\frac{2+\sqrt{2}}{4}\approx 0.854$.
Those asymmetric equilibria at $\delta\neq 1$ are ``bad'' as they
yield lower efficiency than the symmetric equilibrium.  Let $\delta_0$
be the minimum root. When $\alpha\rightarrow 0$, $c\rightarrow
+\infty$, and $\delta_0 = 1/c + o(1/c)\rightarrow 0$.  Then,
$x,y\rightarrow 1$.  Thus, $U\rightarrow 3/2$, $U^\ast\rightarrow 2$,
and $U/U^\ast\rightarrow 0.75$.


\commented{
To summarize, we have that
\begin{lemma}
There are up to three equilibria of the two player, two machine opposite
weight game.  Let $\alpha_0 = \frac{2+\sqrt{2}}{4}$. There are two
cases.
\vspace*{-.5cm}
\begin{itemize}
 \item when $1-\alpha_0\leq \alpha\leq \alpha_0$, there is a unique
 Nash equilibrium, and the efficiency at the Nash equilibrium is at
 least $2\sqrt{2}-2$.  
 \item when $\alpha < 1-\alpha_0$ or $>\alpha_0$, there are three Nash
 equilibria, and the efficiency is at least $0.75$.
\end{itemize}
\end{lemma}
}

From the above simple game, we already observe that the Nash
equilibrium may not be unique, which is different from many congestion
games in which the Nash equilibrium is unique.

For the general two player game, we can show that $0.75$ is actually
the worst efficiency bound with a proof in Appendix~\ref{sec:p2}.
Further, at the asymmetric equilibrium, the utility uniformity
approaches $1/2$ when $\alpha\rightarrow 1$. This is the worst
possible for two player games because as we show in
Section~\ref{sec:results:multi}, a user's utility at any Nash
equilibrium is at least $1/m$ in the $m$-player game. Since the
utility of one player can be at most $1$, and the utility of the other
player be at worst $1/2$, the utility uniformity is at least $1/2$.

Another consequence is that the two player game is always envy-free.
Suppose that the two user's shares are $\vecr_1=(r_{11}, \ldots, r_{1n})$ and
$\vecr_2 = (r_{21}, \ldots, r_{2n})$ respectively.  Then
$U_1(\vecr_1)+U_1(\vecr_2)=U_1(\vecr_1+\vecr_2)= U_1(1,\ldots,1)=1$ because
$r_{i1}+r_{i2}=1$ for all $1\leq i\leq n$.  Again by that $U_1(\vecr_1)\geq
1/2$, we have that $U_1(\vecr_1)\geq U_1(\vecr_2)$, i.e.  any equilibrium
allocation is envy-free.
\begin{theorem}\label{thm:price}
For a two player game, $\pi(Q) \geq 3/4$, $\tau(Q)\geq 0.5$, and
$\rho(Q)=1$.  All the bounds are tight in the worst case.
\end{theorem}

\subsection{Multi-player Game}\label{sec:results:multi}

For large numbers of players, the loss in social welfare can be
unfortunately large.  The following example shows the worst case
bound.  Consider a system with $m=n^2+n$ players and $n$ machines.  Of
the players, there are $n^2$ who have the same weights on all the
machines, i.e. $1/n$ on each machine. The other $n$ players have
weight $1$, each on a different machine and $0$ (or a sufficiently
small $\epsilon$) on all the other machines.  Clearly, $U^\ast=n$.
The following allocation is an equilibrium: the first $n^2$ players
evenly distribute their money among all the machines, the other $n$
player invest all of their money on their respective favorite machine.
Hence, the total money on each machine is $n+1$.  At this equilibrium,
each of the first $n^2$ players receives
$\frac{1}{n}\frac{1/n}{n+1}=\frac{1}{n^2(n+1)}$ on each machine,
resulting in a total utility of $n^3\cdot\frac{1}{n^2(n+1)}<1$.  The
other $n$ players each receives $\frac{1}{n+1}$ on their favorite
machine, resulting in a total utility of $n\cdot \frac{1}{n+1}<1$.
Therefore, the total utility of the equilibrium is $<2$, while the
social optimum is $n=\Theta(\sqrt{m})$.  This bound is the worst
possible. 

What about the utility uniformity of the multi-player allocation
game? We next show that the utility uniformity of the $m$-player
allocation game cannot exceed $m$.

Let $(S_1, \ldots, S_n)$ be the current total bids on the $n$
machines, excluding user $i$. User $i$ can ensure a utility of $1/m$
by distributing his budget proportionally to the current bids. That
is, user $i$, by bidding $s_{ij} = X_i/\sum_{i=1}^{n}S_i$ on machine
$j$, obtains a resource level of:
\[r_{ij} = \frac{s_{ij}}{s_{ij}+S_j} 
    = \frac{S_j / \sum_{i=1}^{n}S_i}{S_j / \sum_{i=1}^{n}S_i+S_j} 
    = \frac{1}{1+\sum_{i=1}^{n}S_i}\,,\]
where $\sum_{j=1}^{n} S_j = \sum_{j=1}^{m} X_j - X_i = m-1$.

Therefore, $r_{ij}=\frac{1}{1+m-1} = \frac{1}{m}$. 
The total utility of user $i$ is
\[\sum_{j=1}^{n}r_{ij}w_{ij} = (1/m)\sum_{j=1}^{n}w_{ij}= 1/m\,.\]

Since each user's utility cannot exceed $1$, the minimal possible
uniformity is $1/m$.

While the utility uniformity can be small, the
envy-freeness, on the other hand, is bounded by a constant of
$2\sqrt{2}-2\approx 0.828$, as shown in~\cite{zhang2004}.  To summarize, we have that
\begin{theorem}
For the $m$-player game $Q$, $\pi(Q)=\Omega(1/\sqrt{m})$, $\tau(Q)\geq
1/m$, and $\rho(Q)\geq 2\sqrt{2}-2$.  All of these bounds are tight in
the worst case.
\end{theorem}

\section{Algorithms}
\label{sec:algorithms}
In the previous section, we present the performance bounds of the game
under the infinite parallelism model.  However, the more interesting
questions in practice are how the equilibrium can be reached and what
is the performance at the Nash equilibrium for the typical distribution of utility
functions.  In particular, we would like to know if the intuitive
strategy of each player constantly re-adjusting his bids according to
the best response algorithm leads to the equilibrium.  To answer these
questions, we resort to simulations. In this section, we present the
algorithms that we use to compute or approximate the best response and
the social optimum in our experiments.  We consider both the infinite
parallelism and finite parallelism model.

\subsection{Infinite Parallelism Model}\label{sec:algorithms:infinite}

As we mentioned before, it is easy to compute the social optimum under
the infinite parallelism model --- we simply assign each machine to
the user who likes it the most. We now present the algorithm for
computing the best response.  Recall that for weights $w_1, \ldots,
w_n$, total bids $y_1,\ldots, y_n$, and the budget $X$, the best
response is to solve the following optimization problem
\[\mbox{maximize $U=\sum_{j=1}^n w_j \frac{x_j}{x_j+y_j}$ subject to}\]
\[\mbox{$\sum_{j=1}^n x_{j} = X$, and $x_{j}\geq 0$.}\]

To compute the best response, we first sort $\frac{w_j}{y_j}$ in 
decreasing order. Without loss of generality, suppose that
\[\frac{w_1}{y_1}\geq \frac{w_2}{y_2}\geq\ldots \frac{w_n}{y_n}\,.\]

Suppose that $\vecx^\ast = (x_1^\ast, \ldots, x_n^\ast)$ is the
optimum solution. We show that if $x_i^\ast=0$, 
then for any $j>i$, $x_j^\ast=0$ too.  Suppose this were not true. Then
\begin{eqnarray*}
\frac{\partial U}{\partial x_j}(\vecx^\ast) & = & w_j\frac{y_j}{(x_j^\ast+y_j)^2} < w_j\frac{y_j}{y_j^2}\\
& = & \frac{w_j}{y_j}\leq\frac{w_i}{y_i}=\frac{\partial U}{\partial x_i}(\vecx^\ast)\,.
\end{eqnarray*}

Thus it contradicts with the optimality condition~(\ref{eqn:opt}).
Suppose that $k = \max\set{i|x_i^\ast>0}$. Again, by the
optimality condition, there exists $\lambda$ such that
$w_i\frac{y_i}{(x_i^\ast+y_i)^2} = \lambda$ for $1\leq i\leq k$,
and $x_i^\ast=0$ for $i>k$.  Equivalently, we have that:
\[x_i^\ast = \sqrt{\frac{w_iy_i}{\lambda}}-y_i\,,\mbox{for $1\leq i\leq k$, and $x_i^\ast=0$ for $i>k$.}\]

Replacing them in the equation $\sum_{i=1}^n x_i^\ast = X$, we can
solve for $\lambda = \frac{(\sum_{i=1}^k\sqrt{w_iy_i})^2}{(X+\sum_{i=1}^k y_i)^2}$. Thus,
\[x_i^\ast=\frac{\sqrt{w_iy_i}}{\sum_{i=1}^k\sqrt{w_iy_i}}(X+\sum_{i=1}^k
y_i)-y_i\,.\]

The remaining question is how to determine $k$? It is the largest value such
that $x_{k}^\ast>0$.  Thus, we obtain the following algorithm to
compute the best response of a user:
\vspace{-.3cm}
\begin{enumerate}
\itemsep = 0pt
\item Sort the machines according to $\frac{w_i}{y_i}$ in decreasing order.
\item Compute the largest $k$ such that
\[
\frac{\sqrt{w_ky_k}}{\sum_{i=1}^{k}\sqrt{w_iy_i}}(X+\sum_{i=1}^{k}y_i)-y_k \geq 0.
\]
\item Set $x_j=0$ for $j>k$, and for $1 \leq j \leq k$, set: 
\[
x_j = \frac{\sqrt{w_jy_j}}{\sum_{i=1}^{k}\sqrt{w_iy_i}}(X+\sum_{i=1}^{k}y_i)-y_j.
\]
\end{enumerate}

The computational complexity of this algorithm is $O(n\log n)$, dominated by the
sorting.  In practice, the best
response can be computed infrequently (e.g. once a minute), so for
a typically powerful modern host, this cost is negligible.

The best response algorithm must send and receive $O(n)$ messages
because each user must obtain the total bids from each host. In practice,
this is more significant than the computational cost. Note that hosts
only reveal to users the sum of the bids on them. As a result, hosts
do not reveal the private preferences and even the individual bids of
one user to another.

\subsection{Finite Parallelism Model}\label{sec:algorithms:finite}

Recall that in the finite parallelism model, each user $i$ only places
bids on at most $k_i$ machines.  Of course, the infinite parallelism
model is just a special case of finite parallelism model in which
$k_i=n$ for all the $i$'s.  In the finite parallelism model, computing
the social optimum is no longer trivial due to bounded parallelism.
It can instead be computed by using the maximum matching algorithm.

Consider the weighted complete bipartite graph $G=(U,V,U\times V)$, where
$U=\set{u_{i\ell}|1\leq i\leq m\,,\mbox{and $1\leq \ell\leq k_i$}}$,
$V=\set{1,2,\ldots, n}$ with edge weight $w_{ij}$ assigned to the 
edge $(u_{i\ell}, v_j)$.  A matching of $G$ is a set of edges with
disjoint nodes, and the weight of a matching is the total weights of
the edges in the matching. As a result, the following lemma holds.
\begin{lemma}
The social optimum is the same as the maximum weight matching of $G$.
\end{lemma}

\up
Thus, we can use the maximum weight matching algorithm to compute the
social optimum.  The maximum weight matching is a classical network
problem and can be solved in polynomial
time~\cite{kuhn1955,fredman1987,gabow1990}.  We choose to implement the
Hungarian algorithm~\cite{kuhn1955,papadimitriou1982} because of its
simplicity. There may exist more efficient algorithm for computing the
maximum matching by exploiting the special structure of $G$.  This
remains an interesting open question.

However, we do not know an efficient algorithm to compute the best
response under the finite parallelism model.  Instead, we provide the
following local search heuristic.

Suppose we again have $n$ machines with weights $w_1, \ldots, w_n$ and
total bids $y_1, \ldots, y_n$. Let the user's budget be $X$ and the
parallelism bound be $k$.  Our goal is to compute an allocation of $X$
to up to $k$ machines to maximize the user's utility.

For a subset of machines $A$, denote by $\vecx(A)$ the best response
on $A$ without parallelism bound and by $U(A)$ the utility obtained by
the best response algorithm.  The local search works as follows:
\vspace{-.3cm}
\begin{enumerate}
\itemsep = 0pt
\item Set $A$ to be the $k$ machines with the highest $w_i/y_i$.
\item Compute $U(A)$ by the best response algorithm on $A$.
\item For each $i\in A$ and each $j\notin A$, repeat
\item \hspace*{.5cm} Let $B = A-\set{i}+\set{j}$, compute $U(B)$.
\item \hspace*{.5cm} If($U(B)>U(A)$), let $A\leftarrow B$, and goto $2$.
\item Output $\vecx(A)$.
\end{enumerate}

\up
Intuitively, by the local search heuristic, we test if we can swap a
machine in $A$ for one not in $A$ to improve the best response
utility. If yes, we swap the machines and repeat the process.
Otherwise, we have reached a local maxima and output that value.  We
suspect that the local maxima that this algorithm finds is also the
global maximum (with respect to an individual user) and that this
process stop after a few number of iterations, but we have not proven
it. However, in our simulations, this algorithm quickly converges to
a high ($\geq .7$) efficiency.

\subsection{Local Greedy Adjustment}\label{sec:algorithms:local}

The above best response algorithms only work for the linear utility
functions described earlier. In practice, utility functions may have
more a complicated form, or even worse, a user may not have a
formulation of his utility function.  We do assume that the user still
has a way to measure his utility, which is the minimum assumption
necessary for any market-based resource allocation mechanism.  In
these situations, users can use a more general strategy, the local
greedy adjustment method, which works as follows. A user finds the two
machines that provide him with the highest and lowest marginal
utility. He then moves a fixed small amount of money from the machine
with low marginal utility to the machine with the higher one.  This
strategy aims to adjust the bids so that the marginal values at each
machine being bid on are the same. This condition guarantees the
allocation is the optimum when the utility function is concave. The
tradeoff for local greedy adjustment is that it takes longer to
stabilize than best-response.

\section{Simulation Results}
\label{sec:simulation_results}

While the analytic results provide us with worst-case analysis for the
infinite parallelism model, in this section we employ simulations to
study the properties of the Nash equilibria in more realistic
scenarios and for the finite parallelism model. First, we determine
whether the user bidding process converges, and if so, what the rate
of convergence is. Second, in cases of convergence, we look at the
performance at equilibrium, using the efficiency and fairness metrics
defined above.

\textbf{Iterative Method.}\hspace*{.2cm} In our simulations, each user
starts with an initial bid vector and then iteratively updates his
bids until a convergence criterion (described below) is met.  The
initial bid is set proportional to the user's weights on the machines.
We experiment with two update methods, the best response methods, as
described in Section~\ref{sec:algorithms:infinite}
and~\ref{sec:algorithms:finite}, and the local greedy adjustment
method, as described in Section~\ref{sec:algorithms:local}.

\begin{figure}[htb]
\centerline{\includegraphics[height=4cm]{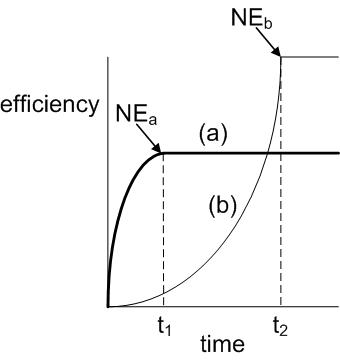}}
\caption{\small Resource allocation scheme (a) converges more slowly
than (b) but achieves an equilibrium with higher efficiency. $NE_a$
and $NE_b$ are the Nash equilibria for (a) and (b), respectively.}
\label{fig:conv_rate_importance}
\end{figure}

\textbf{Convergence Criteria.}\hspace*{.2cm} Convergence time measures
how quickly the system reaches equilibrium.  Convergence time is
particularly important in the context of distributed shared clusters
due to their dynamics. For example, tasks start and finish and users
change their minds about the importance of running tasks. A resource
allocation scheme may result in high efficiency at equilibrium, but
the system may always change before it can reach equilibrium. For
example, of the two resource allocation schemes depicted in
Figure~\ref{fig:conv_rate_importance}, if we expect the dynamics of
the system to change at a rate of approximately $t_1$, we may prefer
resource allocation scheme (a) over (b) even though the efficiency
at $NE_b$ is higher.

There are several different criteria for convergence.  The strongest
criterion is to require that there is only negligible change in the
bids of each user. The problem with this criterion is that it is too
strict: users may see negligible change in their utilities, but
according to this definition the system has not converged. The less
strict \emph{utility gap} criterion requires there to be only
negligible change in the users' utility. Given users' concern for
utility, this is a more natural definition. Indeed, in practice, the
user is probably not willing to re-allocate their bids dramatically
for a small utility gain.  Therefore, we use the utility gap criterion
to measure convergence time for the best response update method, i.e.\
we consider that the system has converged if the utility gap of each
user is smaller than $\epsilon$ ($0.001$ in our experiments).
However, this criterion does not work for the local greedy adjustment
method because users of that method will experience constant
fluctuations in utility as they move money around. For this method, we
use the \emph{marginal utility gap} criterion. We compare the highest
and lowest utility margins on the machines. If the difference is
negligible, then we consider the system to be converged.

In addition to convergence to the equilibrium, we also consider the
criterion from the system provider's view, the \emph{social welfare
stabilization} criterion.  Under this criterion, a system has stabilized
if the change in social welfare is $\leq \epsilon$. Individual users'
utility may not have converged.  This criterion is useful to evaluate
how quickly the system as a whole reaches a particular efficiency
level. 

\textbf{User preferences.}\hspace*{.2cm} We experiment with two models
of user preferences, random distribution and correlated
distribution. With random distribution, users' weights on the
different machines are independently and identically distributed,
according the uniform distribution.  In practice, users' preferences
are probably correlated based on factors like the hosts' location and
the types of applications that users run. To capture these
correlations, we associate with each user and machine a resource
profile vector where each dimension of the vector represents one
resource (e.g., CPU, memory, and network bandwidth).  For a user $i$
with a profile $\vecp_i = (p_{i1}, \ldots, p_{i\ell})$, $p_{ik}$
represents user $i$'s need for resource $k$.  For machine $j$ with
profile $\vecq_j = (q_{j1}, \ldots, q_{j\ell})$, $q_{jk}$ represents
machine $j$'s strength with respect to resource $k$.  Then, $w_{ij}$
is the dot product of user $i$'s and machine $j$'s resource profiles,
i.e.\ $w_{ij} = \vecp_{i} \cdot \vecq_{j} = \sum_{k=1}^\ell
p_{ik}q_{jk}$.  By using these profiles, we compress the parameter
space and introduce correlations between users and machines.

In the following simulations, we fix the number of machines to $100$
and vary the number of users from $5$ to $250$ (but we only report the
results for the range of $5-150$ users since the results remain
similar for a larger number of users).
Sections~\ref{sec:infinite_paral} and~\ref{sec:finite_paral} present
the simulation results when we apply the infinite parallelism and
finite parallelism models, respectively. If the system converges, we
report the number of iterations until convergence. A convergence time of
$200$ iterations indicates non-convergence, in which case we report the
efficiency and fairness values at the point we terminate the
simulation.

\subsection{Infinite parallelism}
\label{sec:infinite_paral}

In this section, we apply the infinite parallelism model, which
assumes that users can use an unlimited number of machines. We present
the efficiency and fairness at the equilibrium, compared to two
baseline allocation methods: social optimum and
weight-proportional, in which users distribute their bids
proportionally to their weights on the machines (which may seem a
reasonable distribution method intuitively).

We present results for the two user preference models. With uniform
preferences, users' weights for the different machines are
independently and identically distributed according to the uniform
distribution, $U\sim(0,1)$ (and are normalized thereafter). In
correlated preferences, each user's and each machine's resource profile vector
has three dimensions, and their values are also taken from the 
uniform distribution, $U\sim(0,1)$.
\begin{figure*}
\begin{tabular}{cc}
\vspace*{-0.8cm}
\epsfig{file=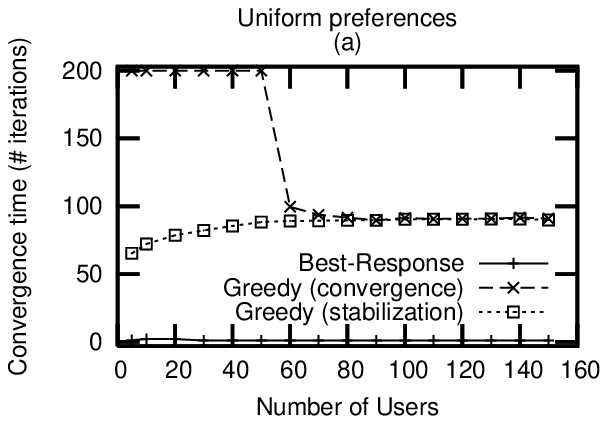,width=2.3in, angle=270} &
\epsfig{file=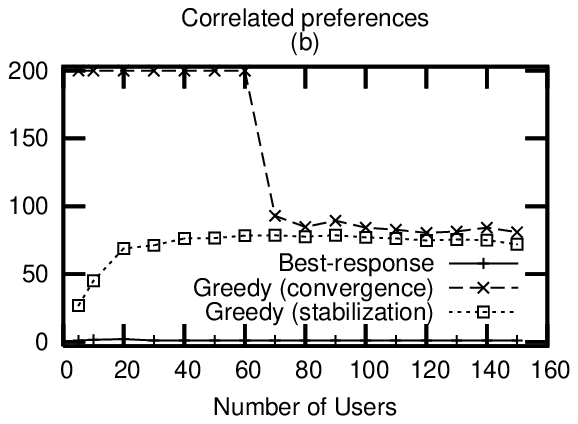,width=2.3in, angle=270}\\
\vspace*{-0.8cm}
\epsfig{file=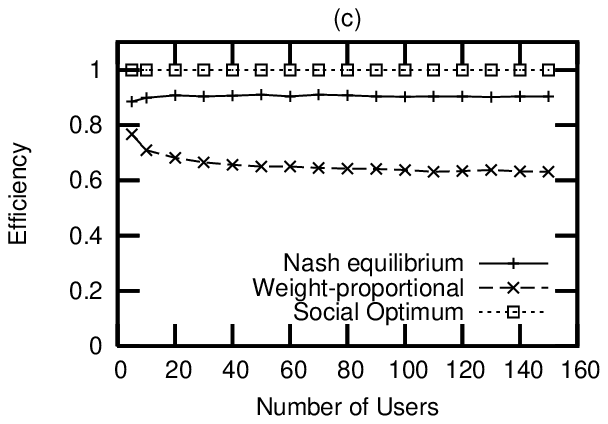,width=2.3in, angle=270} &
\epsfig{file=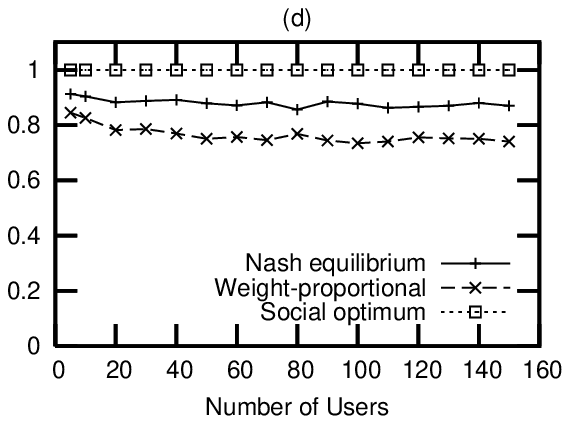,width=2.3in, angle=270}\\ 
\vspace*{-0.8cm}
\epsfig{file=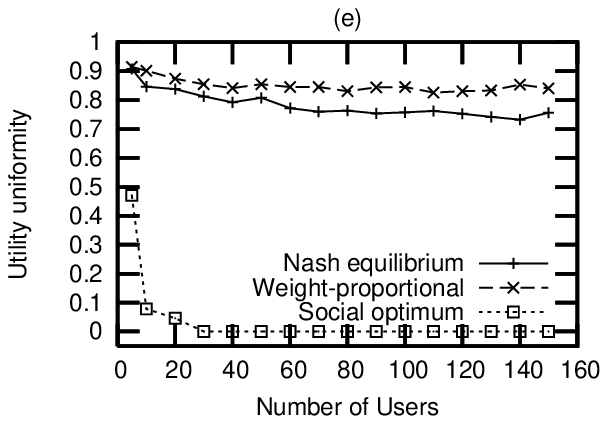,width=2.3in, angle=270} &
\epsfig{file=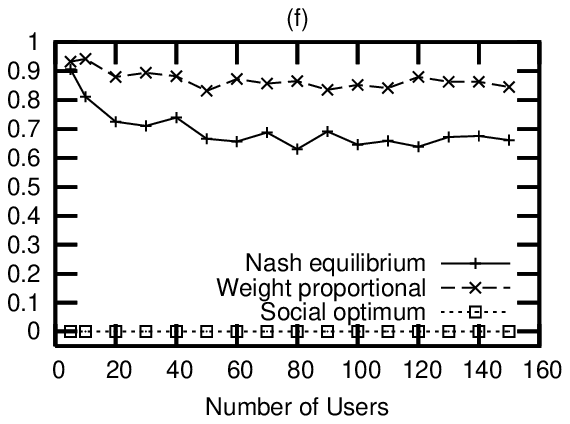,width=2.3in, angle=270} \\
\epsfig{file=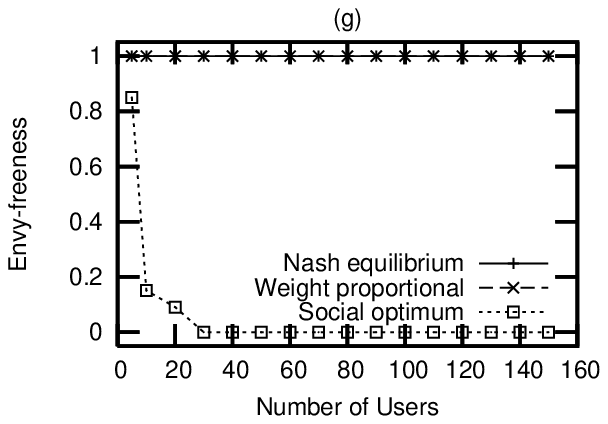,width=2.3in, angle=270} &
\epsfig{file=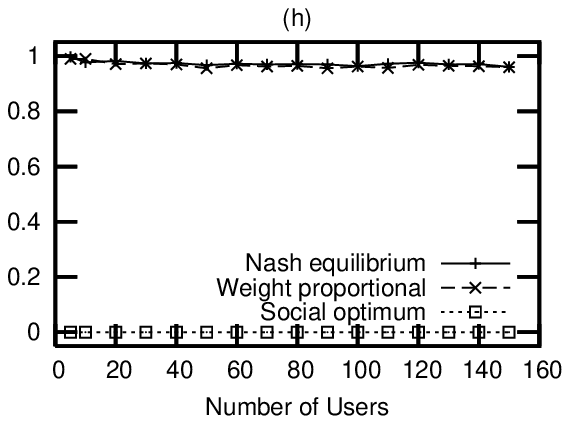,width=2.3in, angle=270}
\end{tabular}
\caption{\small Efficiency, utility uniformity, enviness and
convergence time as a function of the number of users under the
infinite parallelism model, with uniform and correlated
preferences. $n=100$.}
\label{fig:all_metrics}
\end{figure*}

\begin{figure}
\vspace*{-0.4cm}
\epsfig{file=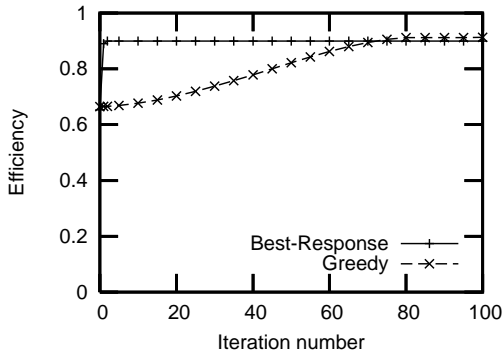,width=2.3in, angle=270}
\vspace*{-0.6cm}
\caption{\small Efficiency level over time under the infinite
parallelism model. number of users = $40$. $n=100$.}
\label{fig:convergence_progress}
\end{figure}

\textbf{Convergence Time.}\hspace*{.2cm} Figure~\ref{fig:all_metrics}
shows the convergence time, efficiency and fairness of the infinite
parallelism model under uniform (left) and correlated (right)
preferences. Plots (a) and (b) show the convergence and stabilization
time of the best-response and local greedy adjustment methods. The
best-response algorithm converges within a few number of iterations
for any number of users. In contrast, the local greedy adjustment
algorithm does not converge even within $500$ iterations when the
number of users is smaller than $60$, but does converge for a larger
number of users. We believe that for small numbers of users, there are
dependency cycles among the users that prevent the system from
converging because one user's decisions affects another user, whose
decisions affect another user, etc. Regardless, the local greedy
adjustment method stabilizes within $100$ iterations.

Figure~\ref{fig:convergence_progress} presents the efficiency over
time for a system with $40$ users. It demonstrates that while both
adjustment methods reach the same social welfare, the best-response
algorithm is faster. 

In the remainder of this paper, we will refer to the (Nash)
equilibrium, independent of the adjustment method used to reach it.

\textbf{Efficiency.}\hspace*{.2cm} Figure~\ref{fig:all_metrics} (c)
and (d) present the efficiency as a function of the number of
users. We present the efficiency at equilibrium, and use the social
optimum and the weight-proportional static allocation methods for
comparison. Social optimum provides an efficient allocation by
definition.  For both user preference models, the efficiency at the
equilibrium is approximately $0.9$, independent of the number of
users, which is only slightly worse than the social optimum. The
efficiency at the equilibrium is $\approx 50\%$ improvement over the
weight-proportional allocation method for uniform preferences, and
$\approx 30\%$ improvement for correlated preferences.

\textbf{Fairness.}\hspace*{.2cm} Figure~\ref{fig:all_metrics}(e) and
(f) present the utility uniformity as a function of the number of
users, and figures (g) and (h) present the envy-freeness. While the
social optimum yields perfect efficiency, it has poor fairness. The
weight-proportional method achieves the highest fairness among the
three allocation methods, but the fairness at the equilibrium is
close. 

The utility uniformity is slightly better at the equilibrium under
uniform preferences ($>0.7$) than under correlated preferences
($>0.6$), since when users' preferences are more aligned, users'
happiness is more likely going to be at the expense of each
other. Although utility uniformity decreases in the number of users,
it remains reasonable even for a large number of users, and flattens
out at some point. At the social optimum, utility uniformity can be
infinitely poor, as some users may be allocated no resources at
all. The same is true with respect to envy-freeness. The difference
between uniform and correlated preferences is best demonstrated in the
social optimum results. When the number of users is small, it may be
possible to satisfy all users to some extent if their preferences are
not aligned, but if they are aligned, even with a very small number of
users, some users get no resources, thus both utility uniformity and
envy-freeness go to zero. As the number of users increases, it becomes
almost impossible to satisfy all users independent of the existence of
correlation.

These results demonstrate the tradeoff between the different
allocation methods. The efficiency at the equilibrium is lower than
the social optimum, but it performs much better with respect to
fairness. The equilibrium allocation is completely envy-free under
uniform preferences and almost envy-free under correlated preferences.

\subsection{Finite parallelism}\label{sec:finite_paral}

\begin{figure}
\vspace*{-0.4cm}
\epsfig{file=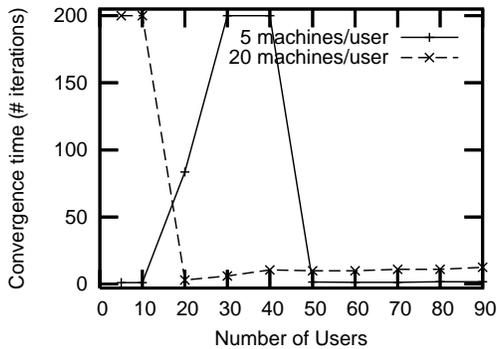,width=2.3in, angle=270}
\vspace*{-0.6cm}
\caption{\small Convergence time under the finite parallelism
model. $n=100$.}
\label{fig:finite_uniform_convergence}
\end{figure}

We also consider the finite parallelism model and use the local search
algorithm, as described in Section~\ref{sec:algorithms:finite}, to
adjust user's bids. We again experimented with both the uniform and
correlated preferences distributions and did not find significant
differences in the results so we present the simulation results
for only the uniform distribution.

In our experiments, the local search algorithm stops quickly --- it
usually only requires two iterations before it discovers a local
maximum.  As we mentioned before, we cannot prove that a local maximum
is the global maximum, but our experiments indicate that the local
search heuristic leads to high efficiency.

\begin{figure}
\vspace*{-0.5cm}
\epsfig{file=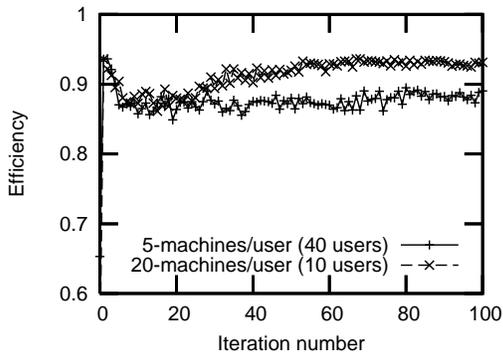,width=2.3in, angle=270}
\vspace*{-0.6cm}
\caption{\small Efficiency level over time under the finite
parallelism model with local search algorithm. $n=100$.}
\label{fig:convergence_progress_finite}
\end{figure}

\textbf{Convergence time.}\hspace*{.2cm} Let $\Delta$ denote the
parallelism bound that limits the maximum number of machines each user
can bid on. We experiment with $\Delta=5$ and $\Delta=20$.  In both
cases, we use $100$ machines and vary the number of users.
Figure~\ref{fig:finite_uniform_convergence} presents the convergence
time for these scenarios.  It shows that the system does not always
converge, but if it does, the convergence happens quickly.  For
$\Delta=5$, the non-convergence occurs when the number of users is
between $20$ and $40$; and for $\Delta=20$, it is when the number of
users is between $5$ and $10$.  In both cases, the ratio of
``competitors'' per machine, $\delta=m\times \Delta/n$ for $m$ users
and $n$ machines, is in the interval $[1,2]$.  We believe that when
this ratio is small, there is no competition in the system so the
equilibrium can be quickly reached. On the other hand, when
competition is high, each machine has bids from many users, so each
user's decision has a small impact on other users, so the system is
more stable and can gradually reach convergence.  However, when there
is ``head-to-head'' competition on each machine, one user's decisions
may cause dramatic changes in another's decisions and cause large
fluctuations in bids.  What is interesting, however, is that although the
system does not converge in these ``bad'' ranges, the system
nontheless achieves and maintains a high level of overall efficiency
after a few iterations (as shown in 
Figure~\ref{fig:convergence_progress_finite}). 
\begin{figure*}
\vspace*{-0.8cm}
\begin{tabular}{cc}
\vspace*{-0.8cm}
\epsfig{file=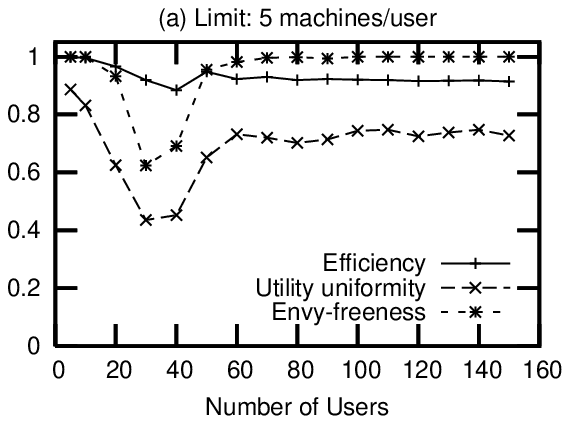,width=2.3in, angle=270}&
\epsfig{file=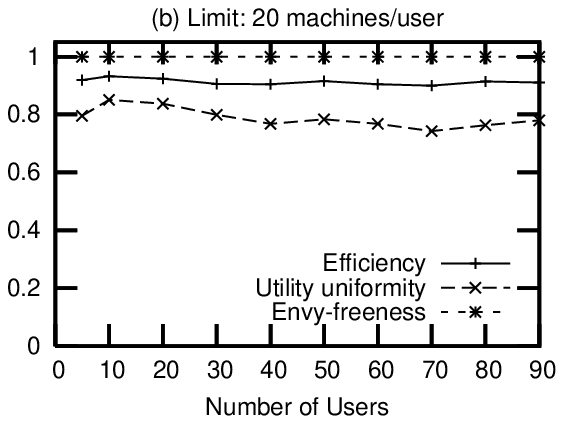,width=2.3in, angle=270}
\end{tabular}
\caption{\small Efficiency, utility uniformity and envy-freeness
under the finite parallelism model. $n=100$.}
\label{fig:finite_uniform}
\end{figure*}

\textbf{Performance.}\hspace*{.2cm} In
Figure~\ref{fig:finite_uniform}, we present the efficiency, utility
uniformity, and envy-freeness at the Nash equilibrium for the finite
parallelism model.  When the system does not converge, we measure
performance by taking the minimum value we observe after running for
many iterations.  When $\Delta=5$, there is a performance drop, in
particular with respect to the fairness metrics, in the range between
$20$ and $40$ users (where it does not converge).  For a larger number
of users, the system converges and achieves a lower level of utility
uniformity, but a high degree of efficiency and envy-freeness, similar
to those under the infinite parallelism model. As described above,
this is due the competition ratio falling into the ``head-to-head''
range. When the parallelism bound is large ($\Delta=20$), the
performance is closer to the infinite parallelism model, and we do not
observe this drop in performance.



\section{Related Work}
\label{sec:relatedwork}

In this section, we describe related work in resource allocation.
There are two main groups: those that incorporate an economic
mechanism, and those that do not.




One class of non-economic algorithms examine resource allocation
from a scheduling perspective (surveyed by Pindedo
\cite{pinedo2002}). One solution is to schedule using FCFS. This
allows an efficient implementation, but does not account for the
differing values of tasks. More sophisticated solutions use
combinatorial optimization (described by Papadimitriou and Steiglitz
\cite{papadimitriou1982}), or examine the resource consumption of
tasks (a recent example is work by Wierman and Harchol-Balter
\cite{wierman2003}). However, these assume that the values and
resource consumption of tasks are reported accurately, which does not
apply in the presence of strategic users. We view scheduling and
resource allocation as two separate functions. Resource allocation
divides a resource among different users while scheduling takes a
given allocation and orders a user's jobs.

Examples of the economic approach are Spawn (by Waldspurger, et
al. \cite{waldspurger1992}), work by Stoica, et
al. \cite{stoica1995}., the Millennium resource allocator (by Chun, et
al. \cite{chun2000}), work by Wellman, et al. \cite{wellman2001},
Bellagio (by AuYoung, et al. \cite{auyoung2004}), and Tycoon (by Lai,
et al. \cite{lai2004-2}). Spawn and the work by Wellman, et al. uses a
reservation abstraction similar to the way airline seats are
allocated. Unfortunately, reservations have a high latency to acquire
resources, unlike the price-anticipating scheme we consider. Bellagio
uses a centralized allocator called SHARE developed by Chun, et
al. \cite{chun2004}. SHARE allocates resources using a centralized
combinatorial auction that allows users to express preferences with
complementarities.  Solving the NP-complete combinatorial auction
problem provides an optimally efficient allocation. The
price-anticipating scheme that we consider does not explicitly operate
on complementarities, thereby possibly losing some efficiency, but it
also avoids the complexity and overhead of combinatorial auctions.

The proportional share abstraction used in the Millennium and Tycoon
resource allocators are the systems which are closest to the fixed
budget, price-anticipating scheme that we examine here. However,
this work focused on implementation and system design issues,
while this work focuses on analysis and simulation.

There have been several analyses~\cite{kelly1997, kelly1998,
hajek2004-2, hajek2004-1, johari2004} of variations of
price-anticipating allocation schemes.  Whereas previous work analyzed
these schemes in the context of allocating network capacity for flows
(with the corresponding topology constraints), we examine allocating
computational capacity for tasks, in which users have task-dependent
private preferences for machines. For example, a user of a scientific
application would prefer machines with faster CPUs and more memory. In
addition, previous work allowed users to have unlimited budgets with a
utility loss for spent funds, while we assume a more realistic limited
budget.  These differences make the game we analyze qualitatively
different from those in previous studies.  For example, there exist
multiple Nash equilibria in our game, and our game does no longer falls
into the category of congestion
games~\cite{rosenthal1973,milchtaich1996} or potential
games~\cite{monderer1996}.

\section{Conclusions}
\label{sec:conclusions}

This work studies the performance of a market-based mechanism for
distributed shared clusters using both analyatical and simulation
methods. We show that despite the worst case bounds, the system can
reach a high performance level at the Nash equilibrium in terms of
both efficiency and fairness metrics. In addition, with a few
exceptions under the finite parallelism model, the system reaches
equilibrium quickly by using the best response algorithm and, when the
number of users is not too small, by the greedy local adjustment
method.

While our work indicates that the price-anticipating scheme may work
well for resource allocation for shared clusters, there are many
interesting directions for furture work.  One direction is to consider
more realistic utility functions. For example, we assume that there is
no parallelization cost, and there is no performance degradation when
multiple users share the same machine.  In practice, both assumptions
may not be correct.  Another assumption is that users have infinite
work, so the more resources they can acquire, the better. In practice,
users have finite work. One approach is address this is to model the
user's utility according to the time to finish a task rather than the
amount of resources he receives. 

Another direction is to study the dynamic properties of the system
when the users' needs change over time, according to some
statistical model.  In addition to the usual questions concerning repeated
games, it would also be important to understand how users should
allocate their budgets wisely over time to accomodate future needs.

\newpage
\bibliographystyle{abbrv}
\bibliography{../bibliographies/resource_allocation,../bibliographies/economics,../bibliographies/networking,../bibliographies/overlay,../bibliographies/network_performance,../bibliographies/peer-to-peer,../bibliographies/security,../bibliographies/reputation,../bibliographies/network_architecture,../bibliographies/grid,../bibliographies/scheduling,../bibliographies/virtualization,../bibliographies/game_theory,../bibliographies/algorithms}

\begin{thebibliography}{10}

\bibitem{planetlab2003}
http://planet-lab.org.

\bibitem{auyoung2004}
A.~AuYoung, B.~N. Chun, A.~C. Snoeren, and A.~Vahdat.
\newblock {Resource Allocation in Federated Distributed Computing
  Infrastructures}.
\newblock In {\em Proceedings of the 1st Workshop on Operating System and
  Architectural Support for the On-demand IT InfraStructure}, 2004.

\bibitem{chun2004}
B.~Chun, C.~Ng, J.~Albrecht, D.~C. Parkes, and A.~Vahdat.
\newblock {Computational Resource Exchanges for Distributed Resource
  Allocation}.
\newblock 2004.

\bibitem{chun2000}
B.~N. Chun and D.~E. Culler.
\newblock {Market-based Proportional Resource Sharing for Clusters}.
\newblock Technical Report CSD-1092, University of California at Berkeley,
  Computer Science Division, January 2000.

\bibitem{ferguson1988}
D.~Ferguson, Y.~Yemimi, and C.~Nikolaou.
\newblock {Microeconomic Algorithms for Load Balancing in Distributed Computer
  Systems}.
\newblock In {\em International Conference on Distributed Computer Systems},
  pages 491--499, 1988.

\bibitem{foster1997}
I.~Foster and C.~Kesselman.
\newblock {Globus: A Metacomputing Infrastructure Toolkit}.
\newblock {\em The International Journal of Supercomputer Applications and High
  Performance Computing}, 11(2):115--128, Summer 1997.

\bibitem{fredman1987}
M.~L. Fredman and R.~E. Tarjan.
\newblock {Fibonacci Heaps and Their Uses in Improved Network Optimization
  Algorithms}.
\newblock {\em Journal of the ACM}, 34(3):596--615, 1987.

\bibitem{gabow1990}
H.~N. Gabow.
\newblock {Data Structures for Weighted Matching and Nearest Common Ancestors
  with Linking}.
\newblock In {\em Proceedings of 1st Annual ACM-SIAM Symposium on Discrete
  algorithms}, pages 434--443, 1990.

\bibitem{hajek2004-2}
B.~Hajek and S.~Yang.
\newblock {Strategic Buyers in a Sum Bid Game for Flat Networks}.
\newblock Manuscript,
  \url{http://tesla.csl.uiuc.edu/~hajek/Papers/HajekYang.pdf}, 2004.

\bibitem{johari2004}
R.~Johari and J.~N. Tsitsiklis.
\newblock {Efficiency Loss in a Network Resource Allocation Game}.
\newblock {\em Mathematics of Operations Research}, 2004.

\bibitem{kelly1997}
F.~P. Kelly.
\newblock {Charging and Rate Control for Elastic Traffic}.
\newblock {\em European Transactions on Telecommunications}, 8:33--37, 1997.

\bibitem{kelly1998}
F.~P. Kelly and A.~K. Maulloo.
\newblock {Rate Control in Communication Networks: Shadow Prices, Proportional
  Fairness and Stability}.
\newblock {\em Operational Research Society}, 49:237--252, 1998.

\bibitem{kuhn1955}
H.~W. Kuhn.
\newblock {The Hungarian Method for the Assignment Problem}.
\newblock {\em Naval Res. Logis. Quart.}, 2:83--97, 1955.

\bibitem{lai2004-2}
K.~Lai, L.~Rasmusson, S.~Sorkin, L.~Zhang, and B.~A. Huberman.
\newblock {Tycoon: an Implemention of a Distributed Market-Based Resource
  Allocation System}.
\newblock Manuscript,
  \url{http://www.hpl.hp.com/research/tycoon/papers_and_presentations}, 2004.

\bibitem{milchtaich1996}
I.~Milchtaich.
\newblock {Congestion Games with Player-Specific Payoff Functions}.
\newblock {\em Games and Economic Behavior}, 13:111--124, 1996.

\bibitem{monderer1996}
D.~Monderer and L.~S. Sharpley.
\newblock {Potential Games}.
\newblock {\em Games and Economic Behavior}, 14:124--143, 1996.

\bibitem{papadimitriou2001}
C.~H. Papadimitriou.
\newblock {Algorithms, Games, and the Internet}.
\newblock In {\em Symposium on Theory of Computing}, 2001.

\bibitem{papadimitriou1982}
C.~H. Papadimitriou and K.~Steiglitz.
\newblock {\em {Combinatorial Optimization}}.
\newblock Dover Publications, Inc., 1982.

\bibitem{pinedo2002}
M.~Pinedo.
\newblock {\em Scheduling}.
\newblock Prentice Hall, 2002.

\bibitem{regev1998}
O.~Regev and N.~Nisan.
\newblock {The Popcorn Market: Online Markets for Computational Resources}.
\newblock In {\em Proceedings of 1st International Conference on Information
  and Computation Economies}, pages 148--157, 1998.

\bibitem{rosen1965}
J.~B. Rosen.
\newblock {Existence and Uniqueness of Equilibrium Points for Concave N-person
  Games}.
\newblock {\em Econometrica}, 33(3):520--534, 1965.

\bibitem{rosenthal1973}
R.~W. Rosenthal.
\newblock {A Class of Games Possessing Pure-Strategy Nash Equilibria}.
\newblock {\em Internation Journal of Game Theory}, 2:65--67, 1973.

\bibitem{hajek2004-1}
S.~Sanghavi and B.~Hajek.
\newblock {Optimal Allocation of a Divisible Good to Strategic Buyers}.
\newblock Manuscript,
  \url{http://tesla.csl.uiuc.edu/~hajek/Papers/OptDivisible.pdf}, 2004.

\bibitem{stoica1995}
I.~Stoica, H.~Abdel-Wahab, and A.~Pothen.
\newblock {A Microeconomic Scheduler for Parallel Computers}.
\newblock In {\em Proceedings of the Workshop on Job Scheduling Strategies for
  Parallel Processing}, pages 122--135, April 1995.

\bibitem{varian1974}
H.~R. Varian.
\newblock {Equity, Envy, and Efficiency}.
\newblock {\em Journal of Economic Theory}, 9:63--91, 1974.

\bibitem{waldspurger1992}
C.~A. Waldspurger, T.~Hogg, B.~A. Huberman, J.~O. Kephart, and W.~S. Stornetta.
\newblock {Spawn: A Distributed Computational Economy}.
\newblock {\em Software Engineering}, 18(2):103--117, 1992.

\bibitem{wellman2001}
M.~P. Wellman, W.~E. Walsh, P.~R. Wurman, and J.~K. MacKie-Mason.
\newblock {Auction Protocols for Decentralized Scheduling}.
\newblock {\em Games and Economic Behavior}, 35:271--303, 2001.

\bibitem{wierman2003}
A.~Wierman and M.~Harchol-Balter.
\newblock {Classifying Scheduling Policies with respect to Unfairness in an
  M/GI/1}.
\newblock In {\em {Proceedings of the ACM SIGMETRICS 2003 Conference on
  Measurement and Modeling of Computer Systems}}, 2003.

\bibitem{zhang2004}
L.~Zhang.
\newblock {On the Efficiency and Fairness of a Fixed Budget Resource Allocation
  Game}.
\newblock Manuscript, 2004.

\end{thebibliography}

\let\e\varepsilon

\appendix

\bigskip

\section{Proof of Theorem 1}\label{sec:p1}
Recall that $w_{ij}$ is the weight of user $i$ on machine $j$, where
$1\leq i\leq m$ and $1\leq j\leq n$.  We consider strongly competitive
game in which for any $1\leq
j\leq n$, there exist $i_1\neq i_2$, such that $w_{i_1j}, w_{i_2j}>0$.
Suppose that $x_{ij}$ is the investment of player $i$ on machine $j$.
Let $Y_j = \sum_{i=1}^m x_{ij}$, the total amount invested on machine
$j$, and $z_{ij} = Y_j - x_{ij}$.  Let $X=\sum_{i=1}^m X_i$, the total
money in the system, and $Z_i = X-X_i$.

Consider the perturbed game $Q^\e$ in which each player's
payoff function is:
\[ U_i^\e (x) = \sum_{j=1}^n w_{ij}\frac{x_{ij}}{\e+Y_j}\,.\]

Or we can think that there is an additional player who invests a tiny
amount of money on each machine so the function $U_i^\e$ is continuous
and concave everywhere.  It is easily verified that $U_i^\e$ is concave
in $x_{ij}$'s, for $1\leq j\leq n$.  The domain of player $i$'s
strategy is the set
\[\Omega_i=\{ (x_{i1},\ldots, x_{in})\,|\,\sum_{j=1}^n x_{ij} = X_i\,,\quad x_{ij}\geq 0\}\,,\]
which is clearly a compact convex set.  Therefore, by Rosen's theorem~\cite{rosen1965},
there exists a Nash equilibrium of the game $Q^\e$.  Pick
$\omega^\e=(x_{ij}^\e)$ be any equilibrium.  Let $\e\rightarrow 0$.
Again, since the strategy space is compact, there exist a infinite
sequence that converge to a limit point.  Suppose the limit point is
$\omega$, i.e. we have a sequence $\e_k\rightarrow 0$ and
$\omega^{\e_k}\rightarrow\omega$.  Clearly, $\omega$ is a legitimate
strategy.  We shall show that $\omega$ is a Nash equilibrium of the
original game $Q=Q^0$.  We can safely assume that for each player $i$,
it has non-zero weights on at least two machines.  The other cases can
be easily handled --- if the weights of a player is all $0$, we can
split his money evenly on all the machines; if a player has only one
positive weight, it will have to invest all of his money on that
machine.  In both cases, they will only increase the reservation price
on some machines and not cause problem to our argument.

Let us consider only those $\e$'s in the converging sequence.  In what
follows, a constant means a number that is solely determined by the
system parameters, $m$, $n$, $w_{ij}$'s, and $X_i$'s, and is
independent of $\e$.  Similarly, let $Y_j^\e = \sum_{i=1}^m
x_{ij}^\e$, and $z_{ij}^\e = Y_j^\e - x_{ij}^\e$.

\begin{lemma}\label{lem:l1}
There exists a constant $M_0, M_1>0$ such that for sufficiently small,
$M_0\leq \lambda_i^\e\leq M_1$ for any $1\leq i\leq m$.
\end{lemma}
\begin{proof}
For any player $i$, it has to invest at least $\frac{X_i}{n}$ on some
machine with positive weight, suppose it is machine $j$.  Then,
\[\lambda_i^\e = w_{ij}\frac{\e+z_{ij}}{(\e+z_{ij}+x_{ij})^2}\,.\]

Therefore, $\lambda_i^\e$ is minimized when $z_{ij}=X-X_i$ and
maximized when $\e+z_{ij}=x_{ij}$.  Thus,
\[\lambda_i^\e\geq w_{ij}\frac{\e+X-X_i}{(\e+X)^2}\,,\]
and
\[\lambda_i^\e\leq \frac{w_{ij}}{4 x_{ij}}\leq \frac{nw_{ij}}{4 X_i}\,.\]

Set $M_0 = \min_{w_{ij}>0}w_{ij}\geq \frac{X-X_i}{4X^2}$ and
$M_1=\max_{w_{ij}>0} \frac{nw_{ij}}{4 X_i}$.  It is easy to verify
that when $\e\leq X$, $M_0\leq \lambda_i^\e \leq M_1$.
\end{proof}

\begin{lemma}\label{lem:l2}
There exists a constant $c_0>0$ such that for sufficiently small
$\e$ and for any $j$, $Y_j^\e \geq c_0$.
\end{lemma}
\begin{proof}
We first show that for sufficiently small $\e$, in the Nash equilbrium
of $Q^\e$, there are at least two players investing on each machine.
For machine $j$, let $w_j, W_j$ be, respectively, the minimum and
maximum nonzero weight on $j$. When $\e< w_j/M_1$ (defined in
Lemma~\ref{lem:l1}), there must be some player investing on machine $j$
because if otherwise, a player $i$'s margin on machine $j$ is
$w_{ij}/\e>M_1$, whenever $w_{ij}>0$.  Thus, there must be some player
investing on player $j$ for sufficiently small $\e$.  Now consider the
situation when there is only one player investing on machine $j$.
Suppose it is player $i$.  Then, $\lambda_i =
w_{ij}\frac{\e}{(\e+x_{ij}^\e)^2}$.  Since $\lambda_i\geq M_0$,
$x_{ij}^\e\leq \sqrt{\frac{w_{ij}\e}{M_0}}-\e$. For another player $k$
with nonzero weight on $j$ (we know there must exist one by
assumption), its margin $\lambda$ on machine $j$ is
\[\frac{w_{kj}}{\e+x_{ij}^\e}\geq w_{kj}\sqrt{\frac{M_0}{w_{ij}\e}}\,.\]
Therefore, when $\e<\frac{w_j^2 M_0}{W_j M_1^2}$, there must be at
least two players investing on $j$.

Now, suppose that there are $k\geq 2$ players investing on machine $j$
in $Q^\e$. Let them be player $1$ to $k$.  Clearly, all of those
players have non-zero weights on $j$.  By Lemma~\ref{lem:l1}, we have
that for any $1\leq i\leq k$, $w_{ij}\frac{\e+z_{ij}^\e}{(\e +
Y_j^\e)^2} \leq M_1$.

Set $M_2 = \max_{w_{ij}>0} M_1/w_{ij}$.  Then,
\[\frac{\e+z_{ij}^\e}{(\e + Y_j^\e)^2} \leq M_1/w_{ij}\leq M_2\,,\quad\mbox{for $1\leq i\leq k$.}\]

Thus,
\begin{eqnarray*}
k M_2 & \geq & \sum_{i=1}^k \frac{\e+z_{ij}^\e}{(\e+Y_j^\e)^2} = \frac{k\e + \sum_{i=1}^k z_{ij}^\e}{(\e+Y_j^\e)^2}\\
& = & \frac{k\e + (k-1)Y_j^\e}{(\e+Y_j^\e)^2} > \frac{k-1}{\e+Y_j^\e}\,.
\end{eqnarray*}

Therefore $Y_j^\e > \frac{k-1}{k M_2}-\e$ for $k\geq 2$.  Set
$c_0=\frac{1}{4M_2}$. When $\e$ is sufficiently small, say
$\leq\frac{1}{4M_2}$, we have that $Y_j^\e \geq c_0$.
\end{proof}

We are now ready for the main lemma,
\begin{lemma}\label{lem:l3}
For any $\delta>0$, for sufficiently small $\e$, we have that
\[\left|\frac{\partial U_i(x)}{\partial x_{ij}}(\omega) - \frac{\partial U_i^\e(x)}{\partial x_{ij}}(\omega^\e)\right| \leq \delta\,.\]
\end{lemma}
\begin{proof}
\begin{eqnarray*}
\frac{\partial U_i(x)}{\partial x_{ij}}(\omega) & = & w_{ij}\frac{z_{ij}}{Y_j^2}\,,\\
\frac{\partial U_i^\e(x)}{\partial x_{ij}}(\omega^\e) & = & w_{ij}\frac{z_{ij}^\e}{(\e+Y_j^\e)^2}\,.
\end{eqnarray*}

The lemma follows immediately by $z_{ij}^\e\rightarrow z_{ij}$ and
$Y_j^\e \rightarrow Y_j$, and that $Y_j^\e\geq c_0$, for some constant
$c_0>0$.
\end{proof}

Now, we are ready to show that $\omega$ is a Nash equilibrium of the
game $Q$.  Suppose it is not true, then the optimum condition is
violated for some player $i$. There are two possibilities.

\begin{enumerate}
\item There are $j,k$, where $j\neq k$, such that $x_{ij},x_{ik}>0$ and
$\frac{\partial U_i}{\partial x_{ij}}\neq\frac{\partial U_i}{\partial
x_{ik}}$.  By Lemma~\ref{lem:l3}, we know that for sufficiently small
$\e$, the following holds: $x_{ij}^\e > 0, x_{ik}^\e > 0$, and
$\frac{\partial U_i^\e}{\partial x_{ij}}(\omega^\e)\neq\frac{\partial
U_i^\e}{\partial x_{ik}}(\omega^\e)$.  This contradicts with that
$\omega^\e$ is a Nash equilibrium of $Q^{\e}$. Now, we assume that
$\lambda_i = w_{ij}\frac{\partial U_i}{\partial x_{ij}}(\omega)$ for
any $x_{ij}>0$.  The other case is then

\item There is $j$ where $x_{ij}=0$ and $w_{ij}\frac{\partial
U_i}{\partial x_{ij}}(\omega) > \lambda_i$.  By the same reason as 1,
we can derive contradiction again.
\end{enumerate}

Hence, $\omega$ is a Nash equilibrium of the game $Q$.\hfill\qed

\section{Proof of Theorem 2}\label{sec:p2}

Suppose that the weights for the two users are $u_1, \ldots, u_n$ and $w_1, \ldots, w_n$, respectively. Denote by $x_1, \ldots, x_n$ and $y_1, \ldots, y_n$ the 
bids for the two users on the machines. We first show that worst case
can always be achieved by when there are only two machines and then show
that the lower bound holds for two users and two machines.

We first make several observations. 
\begin{lemma}\label{lem:m1}
If there is an allocation such that $\frac{\partial U_1}{\partial
x_i}\geq \lambda_1$ and $\frac{\partial U_2}{\partial y_i}\geq \mu_1$,
then there exist a Nash equilibrium with margin $\lambda, \mu$ where
$\lambda\geq \lambda_1$ and $\mu\geq \mu_1$.
\end{lemma}
\begin{proof}
Follows from that $\frac{\partial U_1}{\partial x_i}\geq \lambda_1$ is
a convex set.  We can restrict the strategy to the set $\Omega$
satisfying
\begin{eqnarray*}
u_i\frac{y_i}{(x_i+y_i)^2} & \geq & \lambda_1 \\
w_i\frac{x_i}{(x_i+y_i)^2} & \geq & \mu_1 \\
\sum_{i=1}^ n x_i & = & X\\
\sum_{i=1}^ n y_i & = & Y\\
x_i, y_i & \geq & 0
\end{eqnarray*}

Since $\Omega$ is convex and non-empty, there exists a Nash
equilibrium, say $\omega=(x_1, \ldots, x_n, y_1, \ldots, y_n)$, in the
restricted strategy set $\Omega$.  We show that it is a Nash
equilibrium in the full strategy space.  Otherwise, suppose that
$x=(x_1, \ldots, x_n)$ is not the best response to $y=(y_1, \ldots,
y_n)$.  Then, there are $i,j$ such that
$u_i\frac{y_i}{(x_i+y_i)^2}\neq u_j\frac{y_j}{(x_j+y_j)^2}$.  Let
$u_i\frac{y_i}{(x_i+y_i)^2}>u_j\frac{y_j}{(x_j+y_j)^2}\geq \lambda_0$.
We let $x_i^\ast=x_i+\epsilon$ and $x_j^\ast=x_j-\epsilon$.  Let
$x^\ast$ be the strategy with $x_i$ replaced by $x_i^\ast$ and $x_j$
by $x_j^\ast$. For sufficiently small $\epsilon$, we have that
$U_1(x^\ast, y)>U_2(x, y)$ by
$u_i\frac{y_i}{(x_i+y_i)^2}>u_j\frac{y_j}{(x_j+y_j)^2}$.  Further,
$x^\ast\in \Omega$, contradicting with that $x$ is the best response
to $y$ in $\Omega$.
\end{proof}

Now, we show that if we merge two machines, it will only decrease
the social welfare at the Nash equilibrium.  By merging two machines,
$i$ and $j$, we mean that we replace the two machines with a new machine
with weights $u_i+u_j$ and $w_i+w_j$ for the two users, respectively.
\begin{lemma}\label{lem:m}
Let $Q_1$ be obtained from $Q$ by merging any two machines, then we have
that $U(Q_1)\leq U(Q)$.
\end{lemma}
\begin{proof}
Let $\omega=(x_1, \ldots, x_n, y_1, \ldots, y_n)$ be a Nash equilibrium of
$Q$.  Then,
\begin{eqnarray*}
U_1(\omega) & = & \sum_{j=1}^n u_i \frac{x_i}{x_i+y_i} = \sum_{j=1}^n u_i-u_i\frac{y_i}{x_i+y_i}\\
& = & \sum_{j=1}^n u_i - \sum_{j=1}^n \lambda (x_i+y_i) = 1 - 2\lambda\,.
\end{eqnarray*}

Similarly, $U_2(\omega) = 1-2\mu$.

Now, we show a stronger statement that any Nash equilibrium $\omega$
of $Q$, there exists a Nash equilibrium $\omega_1$ of $Q_1$ such
that
\[ U_i(Q_1, \omega_1)\leq U_i(Q, \omega)\,,\quad\mbox{for any $1\leq i\leq 2$.}\]

We only need to show that there exists $\omega_1$ such that
$\lambda_1\geq \lambda$ and $\mu_1\geq \mu$ where $\lambda_1, \mu_1$
are the respective margin at $\omega_1$ for the two players.  Suppose
that $Q_1$ is obtained by replacing machine $n-1$ and $n$ of $Q$ by
machine $(n-1)'$.  Now consider the Nash equilibrium $\omega$ of $Q$.
Consider the allocation $\omega_0 = (x_1, \ldots, x_{n-2},
x_{n-1}+x_n, y_1,\ldots, y_{n-2}, y_{n-1}+y_n)$.  Let
$\lambda_0,\mu_0$ be the respective margin of the players on machine
$(n-1)'$ at $\omega_0$.  We now show that
$\lambda_0 \geq \lambda$ and $\mu_0\geq \mu$. We wish to show that
\begin{eqnarray*}
&&(u_{n-1}+u_{n})\frac{y_{n-1}+y_n}{(x_{n-1}+x_n+y_{n-1}+y_n)^2}\\
&\geq& u_{n-1}\frac{y_{n-1}}{(x_{n-1}+y_{n-1})^2} = u_{n}\frac{y_{n}}{(x_{n}+y_{n})^2}\,.
\end{eqnarray*}

Let $a_{n-1}=x_{n-1}+y_{n-1}$, $a_n = x_n+y_n$, and $c =
\frac{u_{n-1}}{u_n}$.  Then the above is equivalent to show that
\[\frac{y_{n-1}+y_n}{(a_{n-1}+a_n)^2}\geq
\min \left(c\frac{y_{n-1}}{a_{n-1}^2}, (1-c)\frac{y_{n}}{a_{n}^2}\right)\,.\]

The right hand side achieves maximum at
\[\frac{y_{n-1}y_n}{y_{n-1}a_n^2+y_n a_{n-1}^2}\,,\]
when
\[c=\frac{y_n}{a_n^2}/(\frac{y_{n-1}}{a_{n-1}^2}+\frac{y_{n}}{a_{n}^2})\,.\]

It is now easy to verify that
\[\frac{y_{n-1}+y_n}{(a_{n-1}+a_n)^2}\geq \frac{y_{n-1}y_n}{y_{n-1}a_n^2+y_n a_{n-1}^2}\,.\]

For all the other machines, they still have margin $\lambda$ and
$\mu$.  By Lemma~\ref{lem:m1}, there exists a Nash equilibrium of
$Q_1$ with margin $\lambda_1, \mu_1$ with $\lambda_1\geq \lambda$ and
$\mu_1\geq\mu$.  This proves the lemma.
\end{proof}

By the above lemma, to consider the worst case performance, we divide
the machines into two sets
\[ L = \{ i\,|\, u_i \geq w_i\}\,\quad\mbox{and}\quad  R = \{ i\,|\, u_i < w_i\}\,.\]

If $R=\emptyset$, then $u_i=w_i$ for all the $i$'s. In this case, any
allocation achieves total utility $1$.  So we assume that
$R\neq\emptyset$.

We create a new game $Q_2$ in which there are only two machines $A$
and $B$ where
\[\begin{array}{rclrcl}
u_A & = & \sum_{i\in L} u_i & w_A & = & \sum_{i\in L} w_i \\
u_B & = & \sum_{i\in R} u_i & w_B & = & \sum_{i\in R} w_i
\end{array}\]

By Lemma~\ref{lem:m}, $U(Q_2)\leq U(Q)$. Clearly,
$U^\ast(Q_2)=U^\ast(Q)$ because we allow the player $1$ to occupy
wholly the machine $A$ and player $2$ to occupy $B$, which results a
total utility $u_A+w_B=\sum_{i=1}^n \max(u_i, w_i)$.  Now, our task is
relatively easier, we need to find out the worst performance when
there are only two machines.

Suppose that the two users' weights are, respectively, $(\alpha,
1-\alpha)$, $(\beta, 1-\beta)$ with $0<\alpha,\beta<1$.  And their
allocation is $(x, 1-x)$ and $(y, 1-y)$. Let $s=x+y$ and $\tau =
(2-s)/s$.  We have the following equalities. 
\begin{eqnarray}
s & = & 2/(1+\tau)\label{eqn1}\\
U & = & 2-2(\lambda+\mu)\label{eqn2}\\
\lambda+\mu & = & \frac{\alpha\beta}{s}+\frac{(1-\alpha)(1-\beta)}{2-s}\label{eqn3}
\end{eqnarray}

Equality~(\ref{eqn1}) is obvious.  Equality~(\ref{eqn2}) is proved in
the proof of Lemma~\ref{lem:m}.  Equality~(\ref{eqn3}) is the
consequence of the following equations:
\begin{eqnarray}
\frac{\lambda}{\alpha}+\frac{\mu}{\beta}&=&\frac{1}{s}\label{eqn5}\\
\frac{\lambda}{1-\alpha}+\frac{\mu}{1-\beta}&=&\frac{1}{2-s}\label{eqn6}
\end{eqnarray}

Equation~(\ref{eqn5}) is due to $\alpha\frac{y}{s^2}=\lambda$,
$\beta\frac{x}{s^2}=\mu$, and $s=x+y$.  Clearly, (\ref{eqn3}) follows if
we multiply (\ref{eqn5}) by $\alpha\beta$ and (\ref{eqn6}) by
$(1-\alpha)(1-\beta)$ and then add them.

Now, we define some notations for later use.
\begin{eqnarray*}
b & = & \frac{1}{4\alpha\beta}+\frac{1}{4\beta}+\frac{7}{4\alpha}-2\\
c & = & \frac{1}{2\alpha}+\frac{1}{2\beta}-1\\
d & = & \frac{(1-\alpha)(1-\beta)}{\alpha\beta}\\
e & = & b-c = \frac{5}{4\alpha}+\frac{1}{4\alpha\beta}-\frac{1}{4\beta}-1
\end{eqnarray*}

Since $0<\alpha,\beta<1$, it is easy to verify that $b,c,d,e>0$.

From
\begin{eqnarray*}
\alpha\frac{y}{s^2} & = & (1-\alpha)\frac{1-y}{(2-s)^2}\,,\\
\beta\frac{x}{s^2} & = & (1-\beta)\frac{1-x}{(2-s)^2}\,,\\
\end{eqnarray*}
we have that
\begin{eqnarray*}
\alpha\tau^2 y & = & (1-\alpha)(1-y)\,,\\
\beta\tau^2 x & = & (1-\beta)(1-x)\,.
\end{eqnarray*}

That is, $x=\frac{1}{1+\tau^2(1-\beta)/\beta}$ and
$y=\frac{1}{1+\tau^2(1-\alpha)/\alpha}$. Further,
$x+y=s=\frac{2}{1+\tau}$.  Therefore,
\[\frac{1}{1+\tau^2(1-\beta)/\beta}+\frac{1}{1+\tau^2(1-\alpha)/\alpha}=
\frac{2}{1+\tau}\,.\]

Simplifying the above equation, we obtain that
\begin{equation}
\tau^3 - c\tau^2+c\tau-d = 0\,.\label{eqn4}
\end{equation}

Without loss of generality, we assume that $\alpha\geq \beta, 1-\beta,
1-\alpha$.  Thus, $U^\ast = \alpha+1-\beta$.
Combining
Equalities~(\ref{eqn1},\ref{eqn2},\ref{eqn3}), we have that
\begin{eqnarray*}
U & = & 2-2(\lambda+\mu)\\
 & = & 2-2\left(\frac{\alpha\beta}{s}+\frac{(1-\alpha)(1-\beta)}{2-s}\right)\\
 & = & U^\ast-(\alpha\beta\tau+\frac{(1-\alpha)(1-\beta)}{\tau}+2\alpha\beta-2\beta)\,.
\end{eqnarray*}

Now, we assume that $U/U^\ast < 3/4$ and derive contradiction. If
$U/U^\ast<3/4$, then
\begin{equation}
\alpha\beta\tau+\frac{(1-\alpha)(1-\beta)}{\tau}+2\alpha\beta-2\beta>(1+\alpha-\beta)/4\,.\label{eqnx}
\end{equation}

The above inequality is equivalent to that
$F_{\alpha,\beta}(\tau)>0$, where
\[F_{\alpha,\beta}(\tau) = \tau^2-b\tau+d\,.\]

We now just need to show that for any $0<\beta\leq\alpha<1$,
there does not exist positive root for (\ref{eqn4}) which 
satisfies  $F_{\alpha,\beta}(\tau)>0$. The proof is by algebraic
arguments, and the details are omitted.\hfill\qed



\end{document}